\documentclass[preprint,prd,noshowpacs,nofootinbib]{revtex4}

\usepackage{color}
\usepackage{amssymb}
\usepackage{amsmath}
\usepackage{graphicx}
\usepackage{xcolor}

\usepackage{pdfpages}
\usepackage{epstopdf}
\usepackage[utf8]{inputenc}

\usepackage{mciteplus}
\usepackage{subcaption}
\usepackage{caption}
\usepackage{bbold}
\usepackage{mathbbol}
\DeclareSymbolFontAlphabet{\amsmathbb}{AMSb}%
\usepackage{amsthm}
\usepackage{hyperref}
\usepackage[english]{babel}

\def\be {\begin{equation}}
\def\ee {\end{equation}}
\def\bea {\begin{eqnarray}}
\def\eea {\end{eqnarray}}
\def\bc {\begin{center}}
\def\ec {\end{center}}
\def\bfg {\begin{figure}}
\def\efg {\end{figure}}
\def\bi {\begin{itemize}}
\def\ei {\end{itemize}}

\begin{document}
\title{Unraveling the Mystery of the Cosmological Constant: Does  Spacetime Uncertainty Hold the Key?}

\author{Ahmed~Farag~Ali$^{\triangle \nabla}$}
\email{aali29@essex.edu ; ahmed.ali@fsc.bu.edu.eg}

\author{Nader Inan$^{\Box   \oplus   \otimes }$}
\email{ninan@ucmerced.edu}

\affiliation{\small{$^\triangle$Essex County College, 303 University Ave, Newark, NJ 07102, United States.}}
\affiliation{\small{$^\nabla$Department of Physics, Faculty of Science, Benha University, Benha, 13518, Egypt.}}

\affiliation{$^\Box$ Clovis Community College, 10309 N. Willow, Fresno, CA 93730 USA}
\affiliation{$^\oplus$ University of California, Merced, School of Natural Sciences, P.O. Box 2039,
Merced, CA 95344, USA}
\affiliation{$^\otimes$ Department of Physics, California State University Fresno, Fresno, CA 93740-8031, USA}

\begin{abstract}
\noindent
In addressing the cosmological constant problem, we propose that the discrepancy between the theoretical and observed values can be ascribed to the inherent uncertainty in the spacetime metric. Mach's principle, which posits that mass shapes spacetime, intersects with quantum mechanics' description of a particle as a quantum cloud, rendering the precise location of a particle's mass unknowable. Consequently, understanding spacetime structure at the quantum level becomes elusive. This connection between quantum and spacetime uncertainty could hold the key to resolving the cosmological constant problem. Intriguingly, the length scale of spacetime uncertainty, aligns with the macroscopic quantum weirdness observed in recent experiments. The spacetime uncertainty can be quantified by the scale factor in the Friedmann-Lemaître-Robertson-Walker (FLRW) Universe.

\bigskip
\bigskip
\bigskip

\begin{center}
\noindent
{\em \bf{This essay received an honorable mention in the 2023 Gravity Research Foundation
essay competition.}}
\end{center}

\end{abstract}

\maketitle

\section{Cosmological constant and Uncertainty}

The cosmological constant problem is considered as the most terrible fine-tuning in physics. Weinberg, in his presentation of the cosmological constant problem, used the Planck momentum as the cutoff on the vacuum energy density, and this led to a very big number compared to observations \cite{Weinberg:1988cp}. In this paper, we approach the problem by questioning the Planck scale as a cutoff. Generally, the physical motivation for using the Planck scale as the cut-off is because this is the length scale at which quantum gravity is believed to become relevant. The Planck scale is determined by fundamental constants of nature such as $\hbar$, $G$, and $c$. However, fundamental constants themselves could be varying according to several astrophysical and cosmological studies suggested by Dirac, Barrow, Smolin, Magueijo, Moffat, and others. For example, Dirac proposed that the gravitational constant must be varying with time \cite{Dirac:1975vq,Dirac:1978xh} using a large number hypothesis based on measurements. Several astrophysical and cosmological studies support varying $G$ \cite{Barrow:1996xn,Barrow:1996kc,Barrow:2005hw,Christodoulou:2018xxw,Barrow:2004ad,Moffat:2011rp} that suggest modified theories of gravity. Besides, varying the speed of light with time has been proposed to solve several cosmological puzzles at once \cite{Albrecht:1998ir,Barrow:1999is,Clayton:1998hv}. In addition, the invariance of the Planck constant is constrained by clock correction data from the Global Positioning System \cite{Kentosh:2012zg}. The possibility of a varying fine structure constant has also been explored, most notably by Bekenstein. The fine structure constant, commonly denoted as $\alpha$, is a dimensionless coupling constant that characterizes the strength of the electromagnetic interaction between charged particles. Bekenstein proposed a theoretical framework in which the fine structure constant, $\alpha$, could vary \cite{Bekenstein:1982eu}.  It is clear that if fundamental constants are allowed to vary, there is no meaning to set a unique scale as a Planck scale. Motivated by the possible non-uniqueness of the Planck scale and accounting for the potential variations of fundamental constants in the universe, we introduce another perspective for the cosmological constant problem. We set the cutoff momentum to be {\it unknown} and we intend to determine its value from the limits on our certainty of measurements. Therefore, it is logical to quantify this {\it unknown} cutoff with the {\it uncertainty} of the spacetime metric. In other words, {\it cutoff is due to uncertainty}. To understand this equivalency, let us start with the definition of vacuum energy density as follows.
\begin{eqnarray}
{\rho_{\text{vac}}= \frac{1}{(2\pi \hbar)^3}\int^{P_Z}_0 (\tfrac{1}{2} \hbar \omega_p) ~ d^3p = \frac{1}{16 \pi^3 \hbar^3} \int^{P_{Z}}_0 \sqrt{p^2 c^2+m^2 c^4} ~~ d^3p ,} \label{vacuum0}
\end{eqnarray}
where $P_{Z}$ is the cutoff momentum. We use the standard international units. 

\noindent Notice that Eq. (\ref{vacuum0}) is only applicable in Lorentz geometry and does not account for any curvature of spacetime. It is reasonable to rely on Eq. (\ref{vacuum0}) up to a threshold where the curvature of spacetime begins to have a significant impact. Beyond this cutoff, we no longer have confidence in Eq. (\ref{vacuum0}). This provides a solid basis for our conjecture concerning the equivalency or complementary relationship between the cut-off $P_z$ and the uncertainty of spacetime. To investigate our argument further, let us consider an approximation as $mc <<  p$, the vacuum energy density (in spherical momentum space) reads:
\begin{eqnarray}
{\rho_{\text{vac}}\approx \frac{c}{16 \pi^3\hbar^3} \int^{P_{Z}}_0 p~(4\pi p^2 dp) = \frac{c}{16 \pi^2 \hbar^3} P_Z^4.} \label{vacuum}
\end{eqnarray}
Using de Broglie's duality as follows
\begin{eqnarray}
{P_Z=\frac{\hbar}{L_Z},} \label{deBroglie}
\end{eqnarray}
where $L_Z$ is the cut-off length and can be understood as reduced wavelength of de Broglie's wave \cite{Chang:2011jj}. Eq. (\ref{vacuum}) will be simplified to be
\begin{eqnarray}
\rho_{\text{vac}} \approx \frac{\hbar~c }{16 \pi^2 L_Z^4}
\label{vacuum1}
\end{eqnarray}
The pressing inquiry at this point is, what constitutes the uncertainty in spacetime? The response to this question can be found in multiple studies \cite{Adler:2010wf,Regge:1958wr,Ng:1999hm,Ng:1994zk, Christiansen:2009bz, Mead:1964zz,Vilkovisky:1992pb,dewitt1964gravity,mead1966observable,garay1999quantum}. These studies have calculated the deviation from Lorentz geometry resulting from gravitational fields or curvature as follows \cite{Adler:2010wf}:
\begin{eqnarray}
{\Delta g=\frac{\phi}{c^2}=\frac{G\hbar}{c^3 \l^2}=\frac{{\ell_{Pl}}^2}{l^2}.} \label{curveuncertain}
\end{eqnarray}
In this context, $\Delta g \ll 1$ implies that the geometry is nearly Lorentzian, while $\Delta g = 1$ signifies that spacetime foam starts to have an impact. To arrive at Eq. (\ref{curveuncertain}), Regge, Adler, Jack Ng, and others employed substitutions $E=h\nu=h \frac{c}{l}$ and $\phi= \frac{Gm}{l}$. It is evident that Eq. (\ref{curveuncertain}) establishes an uncertainty in the spacetime metric, as suggested in \cite{Regge:1958wr,Adler:2010wf,Ng:1999hm,Ng:1994zk, Christiansen:2009bz,Mead:1964zz,Vilkovisky:1992pb,dewitt1964gravity,mead1966observable,garay1999quantum}. Now, let's examine Eq. (\ref{vacuum}) and select $L_Z$ as the length scale where uncertainty in Lorentz symmetry begins to appear. This leads to
\begin{eqnarray}
{\Delta g= \frac{{\ell_{Pl}}^2}{L_Z^2}.} \label{uncertainmetric}
\end{eqnarray}
By substituting Eq. (\ref{uncertainmetric}) into Eq. (\ref{vacuum1}), we get
\begin{eqnarray}
{\rho_{\text{vac}}= \frac{\hbar c}{16 \pi^2}\frac{\Delta g^2}{\ell_{Pl}^4} = 10^{74}\Delta g^2~ \frac{\text{(GeV)}^4}{(\hbar c)^3}.} \label{theory}
\end{eqnarray}
The \textit{observed} vacuum energy  found from the cosmological constant, 
$\Lambda \approx 10^{-52}$ m$^{-2}$, is found to be%
\begin{equation}
\rho _{\text{observed}}=\frac{\Lambda c^{4}}{8\pi G}\approx 6\times 10^{-10}~%
\text{J/m}^{3}\qquad \text{or}\qquad \rho _{\text{observed}}\approx 10^{-47}%
\frac{\left( \text{GeV}\right) ^{4}}{\left( \hslash c\right) ^{3}}\label{rho_obs}
\end{equation}
By comparing Eq. (\ref{theory}) with Eq. (\ref{rho_obs}), we find that the uncertainty in the metric should take the following value
\begin{eqnarray}
{10^{74}~ \Delta g^2= 10^{-47}} \implies
{\Delta g= 10^{-61}.}\label{metricuncertain}
\end{eqnarray}
To find the value of $L_Z$ at which uncertainty in spacetime starts to be effective, we find that
\begin{eqnarray}
{L_Z^2=\frac{{\ell_\text{Pl}}^2}{\Delta g}= 2.5\times 10^{-9} \implies
L_Z= 5\times 10^{-5}~m.}\label{L_z}
\end{eqnarray}

\noindent 
The physical meaning of $5\times 10^{-5}~m$ value in \ref{L_z} can be understood from papers by \cite {Freidel:2022ryr, Zeldovich:1968ehl, Tello:2023dqt}. These studies suggest that this value can be found from $L_Z\approx \sqrt{{\ell_{Pl}\ell_{u}}}$ \cite {Freidel:2022ryr}, where $\ell_u$ is the radius of the observable universe. This indicates that $5\times 10^{-5}~m$ is not an arbitrary length scale but is fundamentally related  to the physical parameters of our universe. In fact, it is the geometric mean of the smallest and largest known length scales of  the universe.
Mach's principle \cite{d1899introducing}, harmoniously linked with general relativity, postulates that mass and energy distribution sculpt the fundamental geometry of spacetime \cite{mach1960thescienceofmechanics,Einstein:1916vd}. Nonetheless, quantum uncertainty complicates this idea, as it obstructs our ability to determine the precise location of a particle's mass. Particles are thus depicted as enigmatic clouds, effectively masking their exact mass coordinates. These elusive clouds not only shroud the basic structure of spacetime where the particle navigates but also suggest an incomplete comprehension of spacetime in quantum mechanics due to the inherent limitations of the observation process. This notion of a partial depiction of reality was recognized by Einstein, Podolsky, and Rosen \cite{Einstein:1935rr}. As a result, it can be deduced that the particle cloud epitomizes the intrinsic uncertainty embedded in the spacetime metric. The hydrogen atom serves as an exemplar of this phenomenon, where both the electron and proton lack a clear trajectory \cite{stodolna2013hydrogen}. 

\begin{figure}[htb]
\centering
\includegraphics[width=0.5\textwidth]{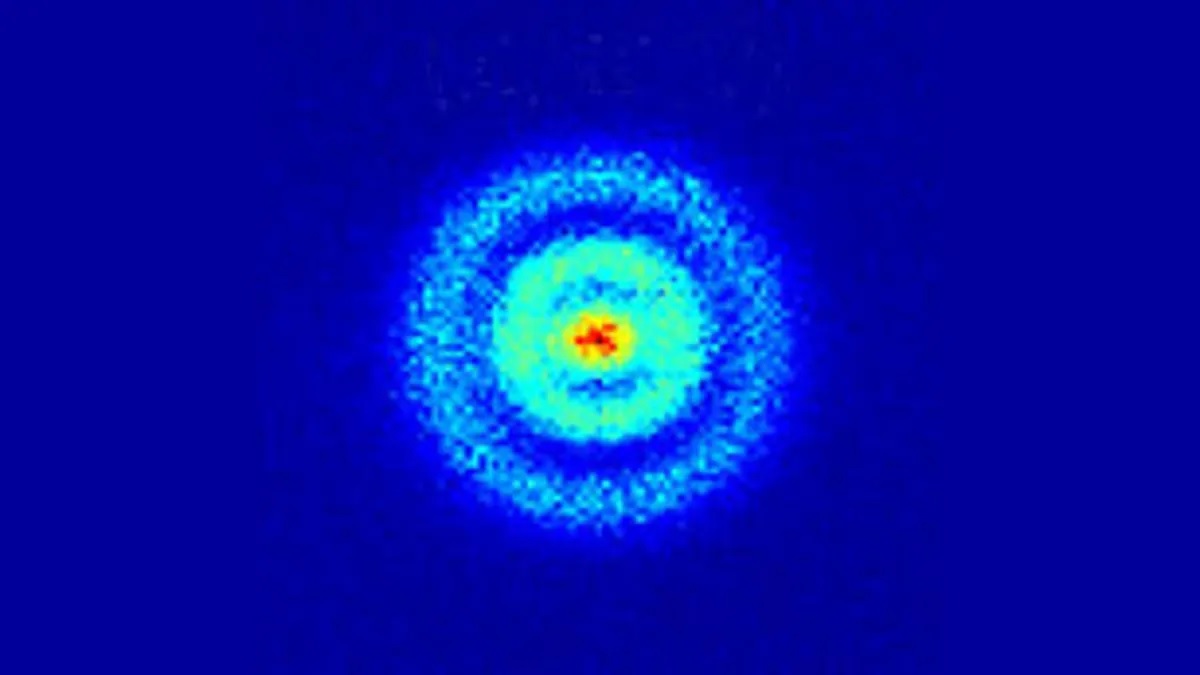}
\caption{Image of Hydrogen atom cloud that can be found in \cite{stodolna2013hydrogen}}
\label{cloud}
\end{figure}
The image of a hydrogen atom cloud (Figure \ref{cloud}) illustrates the constraints that quantum uncertainty imposes on our understanding of spacetime's true nature.

Furthermore, in a groundbreaking study, direct evidence of quantum weirdness was unveiled at a macroscopic scale of approximately $L_{QW}=3\times 10^{-5}$~m by fabricating a quantum drum of that dimension \cite{o2010quantum}. This discovery signifies the largest scale at which quantum weirdness has been detected to date. Interestingly, this scale bears a striking resemblance to the scale $L_Z$, derived from the spacetime metric uncertainty as shown in equation (\ref{L_z}), which we proposed as the cutoff length scale for vacuum energy. This observation posits that the uncertainty principle in quantum mechanics might fundamentally be a manifestation of the spacetime metric's uncertainty, necessitating a nebulous representation for particles instead of well-defined trajectories. In this context, we propose that the close alignment in magnitude order between the cutoff length scale of vacuum energy ($L_Z$) and the macroscopic scale of quantum weirdness ($L_{QW}$) could offer a potential resolution to the cosmological constant problem. In other words, the disagreement between the theoretical and observed values of the cosmological constant might be attributed to the spacetime metric's uncertainty, which gives rise to particle clouds and quantum uncertainty. As such, uncertainty may hold the key to unlocking the solution to the cosmological constant problem.

\section{Scale factor and uncertainty}

\noindent We investigate the potential of attributing uncertainty to the expansion of spacetime. Initially, we can observe that equation \ref{vacuum0} holds true exclusively within Lorentzian geometry. As per empirical findings \cite{slipher1913radial}, the universe is indeed undergoing expansion. Consequently, it is logical to incorporate spacetime curvature in the energy-momentum invariant relationship, denoted by $g_{\mu \nu }p^{\mu}p^{\nu }=-m^{2}c^{2}$. Determining the energy results in%
\begin{equation}
{E=\dfrac{-cg_{0i}p^{i}+c\sqrt{\left( g_{0i}p^{i}\right) ^{2}-g_{00}\left(
g_{ij}p^{i}p^{j}+m^{2}c^{2}\right) }}{g_{00}}.}  \label{energy}
\end{equation}%
which replaces $E=\sqrt{p^{2}c^{2}+m^{2}c^{4}}$ in \ref{vacuum0}. The FLRW universe metric with zero curvature $%
\left( k=0\right) $ reads:%
\begin{equation}
{ds^{2}=-c^{2}dt^{2}+a^{2}\left( t\right) dx^{2},}  \label{FRW}
\end{equation}%
where $a\left( t\right) $ is the cosmological scale factor. Then $\left( \ref{energy}\right) $ becomes%
\begin{equation}
{E=\sqrt{a^{2}\left( t\right) p^{2}c^{2}+m^{2}c^{4}}.} \label{scaleEP}
\end{equation}%
Using equation (\ref{scaleEP}) in \ref{vacuum0} and letting $mc<<p$ gives%
\begin{equation}
{\rho _{\text{vac}}=\dfrac{c}{\left( 2\pi h\right) ^{3}}%
\int_{0}^{P_{Z}} a\left( t\right) p~d^{3}p.} \label{rho_vac}
\end{equation}%
The solution for $a\left( t\right) $ for a universe dominated by dark energy
is known to be  $a\left( t\right) =a_{0}e^{\pm 2\lambda t}$
where $\lambda \equiv H\left( t_{0}\right) \sqrt{\Omega _{\Lambda }\left(
t_{0}\right) }$, $H=a^{-1} da/dt$ is the Hubble parameter, and $t_{0}$ is the current epoch of the universe. Note that 
$\Omega _{\Lambda }\left( t_{0}\right) =\rho \left( t_{0}\right) /\rho _{c}$%
, where $\rho _{c}$ is the critical mass density. Since $\rho \left(
t_{0}\right) \approx \rho _{c}$ in the current epoch, then $\lambda \approx
H\left( t_{0}\right) \equiv H_{0}$. Then $\left( \ref{rho_vac}\right) $ becomes%
\begin{equation}
{\rho _{\text{vac}}=\dfrac{ca_{0}e^{\pm 2\lambda t}}{4\pi ^{2}\hslash ^{3}}%
\int_{0}^{P_{Z}}p^{3}dp=\dfrac{ca_{0}e^{\pm 2\lambda t}}{16\pi ^{2}\hslash
^{3}}P_{Z}^{4}.} \label{geometricrho}
\end{equation}%
We use equation (\ref{deBroglie}) and 
 evaluate the expression for $t=T_{Z}$ which is the appropriate cut-off 
\textit{time} scale for which $\left( \ref{rho_vac}\right) $ is valid.
Normalizing to $a_{0}=1$ and using $P_{Z}=\hbar/L_{Z}$ leads to
\begin{equation}
{T_{Z}\sim \dfrac{1}{2H_{0}}\ln \left( \dfrac{16\pi ^{2}L_{Z}^{4}\rho _{obs}}{%
\hslash c}\right) \sim 10^{9}yr.}
\end{equation}%
where we have used $L_Z= 5\times 10^{-5}$ m since this was shown to be the appropriate cut-off length scale in \ref{L_z}. Therefore, we find that the appropriate cut-off time scale is approximately the age of the universe, rather than the Planck time, $t_{Pl}\sim 10^{-44}s$, which is
implied by the usual choice of using a cut-off momentum, $P_{Pl}=h/L_{Pl}$ where $L _{Pl}=ct_{Pl}$.


Now consider the relationship between coordinate length $\left( L_{C}\right) 
$ and proper length $\left( L_{\text{proper}}\right) $ for the metric in $\left( %
\ref{FRW}\right) $,%
\begin{equation}
{L_{\text{proper}}=-\int_{0}^{L_{C}}\sqrt{g_{\mu \nu }dx^{\mu }dx^{\nu }}%
=-\int_{0}^{L_{C}}\sqrt{-c^{2}dt^{2}+a^{2}\left( t\right) dx^{2}}=-c\int_{0}^{T}%
\sqrt{-1+a^{2}\left( t\right) v^{2}/c^{2}}~dt,}  \label{L_proper}
\end{equation}%
where we have integrated from the start of the universe to some time $T$
associated with the coordinate length $L_{C}$. For approximation, we may consider slow observer velocities, $\sqrt{-1+a^{2}\left( t\right) v^{2}/c^{2}}\approx -1+\tfrac{1}{2c^{2}}%
a^{2}\left( t\right) v^{2}$. Then $\left( \ref{L_proper}\right) $
becomes%
\begin{equation}
L_{\text{proper}}=cT-\tfrac{a_{0}^{2}v^{2}}{8c\lambda }\left(e^{4\lambda
T}-1\right) = c T- \frac{\ell_{Pl}}{\Delta g^{1/2}}
\end{equation}%
Here we have identified the deviation of the proper length from Lorentzian geometry (the term involving the scale factor) as being associated with the metric uncertainty. Therefore, normalizing to $a_{0}=1$ leads to

\begin{equation}
\Delta g^{1/2}=\dfrac{8c\ell_{Pl}\lambda}{v^{2}\left(e^{4\lambda T}-1\right)}\label{delta_g}
\end{equation}%
which gives the uncertainty of the spacetime metric in the case of the FLRW universe. This result demonstrates that uncertainty in the spacetime metric exponentially decreases with time which means that as the universe expands, it behaves more and more classical as the spacetime uncertainty diminishes. However, the current non-zero spacetime uncertainty in $\left( \ref{metricuncertain}\right)$ could be the root cause for the seeming discrepancy between the vacuum energy determined by observing the expansion of the universe, and the vacuum energy determined by evaluating the sum of all quantum vacuum modes. Note that Eq $\left( \ref{delta_g}\right)$ implies that $\lambda$ is time-dependent which has been investigated in some approaches utilizing a time-varying cosmological constant \cite {Peebles:1987ek}.

\section{conclusion}
Mach's principle, in harmony with general relativity, asserts that mass and energy distribution shape the fundamental geometry of spacetime, yet quantum uncertainty complicates this notion by hindering our ability to pinpoint a particle's precise mass location. Consequently, particles are depicted as enigmatic clouds obscuring their exact mass coordinates, veiling the underlying structure of spacetime and suggesting an incomplete understanding of space within quantum mechanics due to inherent observation limitations. This inherent uncertainty in the spacetime metric could be linked to the discrepancy between the theoretical and observed values of the cosmological constant, thereby contributing to the emergence of particle clouds and quantum uncertainty. Exploring this relationship may yield vital insights into the cosmological constant problem and enrich our comprehension of the interplay between quantum mechanics and spacetime. This proposes that the macroscopic behavior of the universe is intrinsically connected to the microscopic world of quantum mechanics.

\section*{Acknowledgment}

The authors are grateful to Douglas Singleton for  valuable discussions.

\bibliographystyle{utcaps}
\bibliography{ref.bib}{}







\end{document}